# Charge Mobility and Dynamics in Spin-crossover Nanoparticles studied by Time-Resolved Microwave Conductivity

†‡Julien Dugay*, †§Wiel Evers, ‡Ramón Torres-Cavanillas, ‡Mónica Giménez-Marqués, ‡Eugenio Coronado* and †Herre S.J. Van der Zant.

† Kavli Institute of Nanoscience, Delft University of Technology, Lorentzweg 1, 2628 CJ Delft, The Netherlands

‡Instituto de Ciencia Molecular (ICMol), Universidad de Valencia, Catedrático José Beltrán 2, 46980 Paterna, Spain

§Opto-Electronic Materials Section, Delft ChemTech, Faculty of Applied Sciences, Delft University of Technology, Julianalaan 136, 2628 BL Delft, The Netherlands

* Corresponding authors: J.D (Julien.dugay@gmail.com) and E.C (eugenio.coronado@uv.es)

**ABSTRACT:** We use the electrode-less time-resolved microwave conductivity (TRMC) technique to characterize spin-crossover (SCO) nanoparticles. We show that TRMC is a simple and accurate mean for simultaneously assessing the magnetic state of SCO compounds and charge transport information on the nanometre length scale. In the low-spin state from liquid nitrogen temperature up to 360 K the TRMC measurements present two well-defined regimes in the mobility and in the half-life times, possessing similar transition temperatures $T_R$ near 225 K. Below $T_R$, an activation-less regime associated with short lifetimes of the charge carriers points at the presence of shallow-trap states. Above $T_R$, these states are thermally released yielding a thermally activated hopping regime where longer hops increases the mobility and, concomitantly, the barrier energy. The activation energy could originate from intricate contributions such as polaronic self-localizations, but also from dynamic disorder due to phonons and/or thermal fluctuations of SCO moieties.

## INTRODUCTION

Spin-crossover (SCO) materials form a remarkable class of compounds with the ability to reversibly change their ground spin state between a low-spin (LS) and a high-spin (HS), triggered by a rich palette of external stimuli.[1–4] Temperature is the most common stimulus used to promote the entropy-driven population of the HS state at elevated temperatures subverting the physical properties of the SCO materials such as a change in their magnetic moment, structure and colour. Moreover, first-order phase-transitions arise when elastic interactions within dense arrays of SCO molecules overcome a certain threshold. When even stronger elastic interactions take place, thermal hysteresis loops may appear spanning room temperature This feature may be of interes[2,3]t for developing future technological applications.[1,5] In particular, powders of nanoparticles (NPs) as small as 4 nm made of the 1D polymeric $[Fe(Htrz)_2(trz)](BF_4)$ SCO compound show abrupt transitions above room temperature maintaining wide thermal memory effects of *ca.* 25 K.[6] Moreover, it has been previously shown that the amount of light reflected as a function of temperature is a simple and accurate mean to assess the magnetic state of the studied SCO compounds.[7,8] Furthermore, these memory effects have been probed by electrical conductivity measurements both in powdered samples,[9–11] but also in micro-[12,13,14,15] and nano-structures[16,17,18] down to the single NP level.[19]

When incorporated into electronic devices, the conduction of the NP increases or decreases upon changing the ground spin state. From the LS to HS state, some of us showed that single NP devices exhibit an increase of the conductance,[19] while the reverse situation occurs for small assembles of similar SCO NPs.[17] A consistent decrease of the conductance in the HS state has also been demonstrated by the Bousseksou's group while investigating larger assemblies of *ca.* 15 nm sized NPs[20] and micro-rods[14,15,20–22] as well as in powdered-samples of the same compound.[9,11] At this stage, it is tempting to attribute these opposite behaviours to separate charge transport regimes: *i)* single-electron tunnelling transport within a double barrier tunnelling configuration for single NPs, when the gap size of the electrodes is of *ca.* 5-10 nm, as evidenced by Coulomb staircase in the current-voltage characteristics at 10 K and the absence of thermal activation of the conductance above room temperature[19] and *ii)* hopping transport for NP assemblies placed on gap sizes ranging from *ca.* 50 nm[17] up to several microns.[9,11,14,15,20–22] For larger SCO objects and electrode separation, a charge transport dominated by polaron hopping was proposed by the Bousseksou's group, based on the thermally activated conductivity above room temperature combined with rather small conductivity values.[9]

In the previous studies charge carriers were generated via source-drain electrodes in direct contact with the SCO compounds. The influence of the electrodes and/or interparticle transport can be significantly reduced using AC electric fields as it confines the motion of the charge carriers into a smaller spatial area. Recently, thermal hysteresis loops in compacted powder samples of SCO microrods have been measured by broadband ($10^{-2}$ - $10^6$ Hz) AC conductivity experiments, but also from the dielectric losses and relaxation dynamics of the charge carriers.[11,22] However, the single-electron tunnelling to hopping transition aforementioned could not be proved in these cases, which requires higher frequencies: AC transport measurements should then be carried out at a greater electric field frequency to reach a charge displacement in the nanometre range.

Time-resolved microwave conductivity (TRMC) is a powerful method to quantify charge carrier mobilities and dynamics for various systems such as organic and inorganic semiconducting materials and photoactive molecular layers.[23,24] TRMC measurements do not employ ohmic contacts, thereby eliminating perturbations related to grain boundaries, dielectric polarization effects, space-charge effects, contact resistances, differences in the NP coupling and charge injection at electrodes.

In this work, we report on the first TRMC study applied to SCO compounds. The mobilities of both, electrons and holes as well as their relaxation kinetics have been monitored by measuring the absorbed power of a 9 GHz microwave field as a function of time.

## RESULTS

Our investigations are focused on 100 nm size hybrid core-shell $SCO@SiO_2$ NPs, based on the SCO coordination polymer $[Fe(Htrz)_2(trz)](BF_4)$ stabilized by an inorganic shell of silica (see supporting material for more details). This hybrid system was selected in view of its enhanced robustness when submitted to thermal cycles in the range 300-400 K with the aim of measuring its hysteretic behavior.[18]

### Characterization of the $SCO@SiO_2$ NPs.

**Magnetic properties:** The temperature-dependent magnetic susceptibility data of a powder sample of $SCO@SiO_2$ NPs is presented in Fig. 1. Magnetic data were collected using a SQUID magnetometer and correspond to the second temperature cycle performed with heating and cooling constant sweep rates of 1 K·min$^{-1}$ under a magnetic field of 0.1 T. The spin transition features, regarding the value in the HS state of ~3.5 cm$^3$·K·mol$^{-1}$, the steepness of ~10 K and the 40 K-wide thermal hysteresis loop, are consistent to that of the previous reported on similar systems.[25]

**Composition:** A representative high-angle annular dark-field scanning transmission electron microscopy HAADF-STEM image of the SCO@SiO$_2$ NPs is shown in the inset of **Figure 1** (see also the supplementary **Figure S1)**. These NPs exhibit a rod-like shape with an average length of 109 ± 24 nm and a diameter of 45 ± 30 nm; energy-dispersive X-ray (EDX) experiments confirmed the presence of an outer pure silica shell surrounding the NP cores as shown in **Figure S2, Supporting Information**.

**Optical properties:** The total fraction of light absorbed ($F_A$) by a powder sample of SCO NPs compared to a highly reflecting reference sample at a fixed wavelength was estimated following Eq.1:

$$F_A = 1 - (F_R + F_T), \quad (1)$$

where $F_T$, the incident laser light fraction transmitted through the powder sample, is equal to zero (see **Fig. S4**). Optical diffuse reflectance ($F_R$) from powder samples of SCO@SiO$_2$ NPs have been acquired using a CARY 5G UV-Vis-NIR spectrophotometer adapted with an integrating sphere, in the wavelength range 300-800 nm. The SCO core of the NPs exhibited well-known features of Iron(II) SCO complexes, that is to say, the absorption band in the visible centred at 530 nm (see **Fig. S5**). Accordingly, this absorption band corresponds to one of the two spin-allowed ligand-field transitions $^1A_1$ - $^1T_1$ in the LS state. The value of $F_A$ was extracted in the low-spin state from the absorption spectra at 530 nm ($F_A$ = 69 %). These values have been exploited to quantitatively estimate the mobilities of SCO NPs measured via TRMC (see Eq.2).

## Time-resolved microwave conductance (TRMC) measurements.

TRMC is a contactless technique that utilizes high-frequency microwaves to probe the conductivity increase induced by a laser pulse, an electron beam or X-rays. In this work, the charge carriers are photogenerated using an accordable laser and the transport properties are probed using 9 GHz microwaves (for more details see Supplementary Information and **Fig. S6**). The high-frequency microwave signal is provided by a peak amplitude of *ca.* 100 V/cm, which is low enough to prevent alteration of the compound itself and charge diffusion after their photogeneration. TRMC techniques offer a local probing scale typically close to the polymer chain or intermolecular distance.[24,23]

### Effect of dose in the pulse on charge carrier mobility

A quartz sample holder was specially designed to perform the time-resolved microwave conductivity measurements of powder samples of SCO@SiO$_2$ NPs (see an optical image taken at 300 K in **Fig. 2** and versus temperature in **Movie S1**). As the width of the quartz holder is thinner than the powder thickness used for reflectance measurements, we ensured that $F_T$ remained close to zero when the SCO NP powder was present in the quartz holder using diode laser excitation at 530 nm and a spectrophotometer in the wavelength range 300-900 nm (see **Fig. S4**, Supporting Information).

**Figure 2** presents the maximum TRMC signal change as a function of incident photon density per laser pulse, $J_0$. The latter was varied at room temperature using metallic neutral density filters. The photon wavelength was chosen to fall close to the peak of the LS state absorbtion band of the SCO NPs, *i.e.*, λ = 530 nm. The TRMC signal is expressed as the product of the optical charge carrier generation quantum yield, ϕ, and the sum of electron and hole mobility, $\Sigma\mu$. This product is calculated according to:[24]

$$\phi . \Sigma\mu = \frac{\Delta G_{max}}{\beta . e . I_0 . F_A}, \quad (2)$$

where $\Delta G_{max}$ is the maximum measured change in conductance, $\beta$ is the ratio between the quartz holder height and width perpendicular to the microwave vector, $e$ is the elementary charge, $I_0$ is the number of photons per unit area per pulse, and $F_A$ is the fraction of absorbed light.

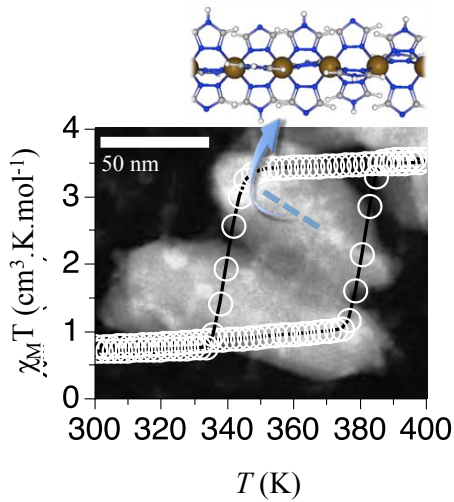

**Fig. 1** Magnetic thermal hysteresis loop of a powder sample of hybrid SCO NPs, based on the coordination polymer [Fe(Htrz)$_2$(trz)](BF$_4$). This system exhibits a wide and abrupt hysteresis loop above room temperature. Inset: a representative TEM image acquired in STEM-HAADF mode of these NPs. The highly scattering areas appear as bright regions.

The maximum conductivity change $\phi \cdot \Sigma\mu$, as pointed out by the red arrow in inset of this figure, was extracted from 1000 averaged spectra after laser-pulse excitation and systematically subtracted from the dark conductance of the sample. The net yield of mobile charge carriers per absorbed photon $\phi$ is not known for SCO compounds, however, it has to be smaller or equal to 1. At short time scales after the pulse, $\phi$ is assumed to be close to unity, *i.e.*, $\phi = \phi_{max} = 1$.[26] The product $\phi_{max} \cdot \Sigma\mu$ corresponds therefore either to a good approximation of the genuine electrical mobility of the charge carrier values or to a lower limit, referred after to as $\Sigma\mu_{min}$. A charge carrier mobility of $\sim 1.2 \times 10^{-5}$ cm$^{-2}$ / (V.s) is obtained at $2.4 \times 10^{14}$ photons/cm² per pulse, which is the lowest intensity that still showed a sufficient signal-to-noise ratio. On the other hand, we observed a smaller mobility value of $\sim 5.8 \times 10^{-6}$ cm$^{-2}$/(V.s) for $2.45 \times 10^{15}$ photons/cm² per pulse, which is the highest intensity delivered by the TRMC setup.

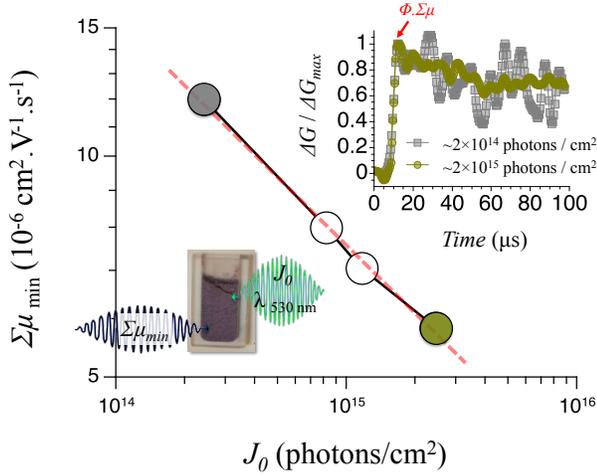

**Fig. 2** Maximum observed TRMC signals measured at room temperature as a function of incident photons per laser pulse $J_0$ excited at $\lambda = 530$ nm. The inset shows two TRMC signals (corresponding to the grey and yellow circles), normalized to the peak signal for clarity.

The behaviour of the maximum conductivity against the incident intensity is commonly described via a power law ($\Sigma\mu_{min} \propto J_0^{-\delta}$).[27] In the case where only first order decay processes take place (*i.e*, $\delta = 1$), higher-order recombination occurs within the nanosecond laser when $\delta < 1$. In **Figure 2**, $\delta$ equals 0.34 as deduced from the linear fit (red dashed line), suggesting that higher-order processes indeed occur at room temperature, at least for a decade of intensity change.[28] For all the next experiments, we fixed the laser intensity to the highest value of $2.45 \times 10^{15}$ photons/cm² per pulse offering the best signal-to-noise ratio.

## High-temperature dependence of the charge carrier mobility using TRMC.

**Figure 3** presents the extracted maximum values of the conductivity spectra for temperatures in the range of $300 < T < 400$ K. Representative microwaves conductivity spectra as a function of time in the heating and cooling modes at an elevated temperature of 350 K are shown in **Fig. S7,** Supplementary Information. Each spectrum was recorded after several tens of minutes to ensure thermal equilibrium and averaged over 1000 laser pulses. A clear difference in the change of conductance values can be observed depending on the thermal history of the sample depicting a 25 K thermal hysteresis loop. We would like to stress that this hysteresis effect was reproducible over several weeks and that an empty quartz sample holder did not show such a temperature dependence as can be seen in **Fig. S7**. Moreover, after all the TRMC measurements were taken, the same powder sample exhibited a well-known thermochromic effect above room temperature from purple to white/yellow, triggered by means of a hot plate (see also **Movie S1**). Hence, we can ascribe the hysteresis loop in the change of conductance to the associated first-order spin transition of these hybrid SCO@SiO$_2$ NPs. The origin of the thermal hysteresis loop in the change in conductance can be understood through the thermal reflectivity change of our SCO compounds, to which the TRMC technique is very sensitive. The amount of light reflected as a function of temperature has previously been proved to be a simple and accurate mean to assess the magnetic state of the studied SCO compounds.[7,8] An advantage given by the TRMC in this work is the ability to simultaneously obtain the thermal hysteresis loop features of SCO NPs and quantitative values of mobilities and dynamics of the charge carriers for different powder samples.

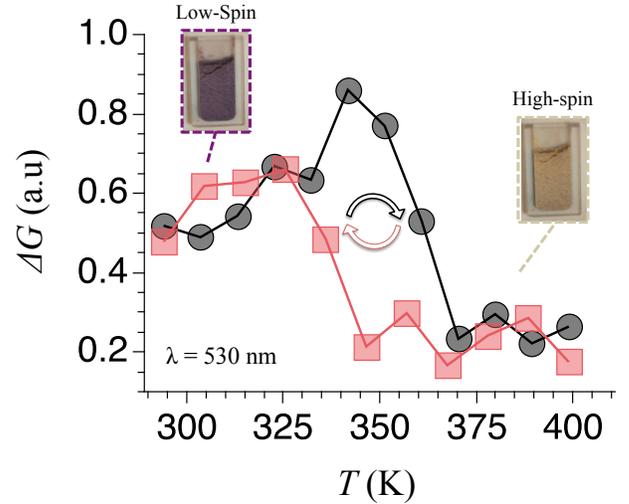

**Fig. 3** Normalized TRMC signals against temperature from room temperature up to 400 K. Optical photographs of the measured powder sample compacted in a quartz holder are shown on the left and right in the LS and HS states, respectively. The signal drop coresponds to an expected decrease of absorbance at $\lambda = 530$ nm, to which the TRMC technique is sensitive.

Mobility values in the LS state are in the range of ~ $1\times10^{-(6\ \text{to}\ 5)}$ cm$^2$.V$^{-1}$.s$^{-1}$, in good agreement with the experimental value reported for a comparable SCO compound based on Fe(II). For instance, thin films of Fe(phen)$_2$(NCS)$_2$ showed a hole mobility of $6.53\times10^{-6}$ cm$^2$.V$^{-1}$.s$^{-1}$ at room temperature (extracted from space charge-limited regime).[29] Nevertheless, the nature of the charge carriers remains unknown for the triazole-based SCO NPs as both electrons and holes can contribute to the observed photoconductivity using TRMC. It is also worthwhile to note that the extracted mobilities stand for the ensemble of NPs, which is isotropic as they are randomly oriented. Therefore, if it turns out that the transport is 1D, which is not yet proved for this SCO compound, the mobility that we measured in this work should be multiplied by a factor of 3.[30]

In an attempt to quantitatively compare the transport properties of the LS state with those in HS state, we photogenerated charge carriers at a longer wavelength centred at λ = 830 nm corresponding to HS optical absorption band maximum. As shown in **Fig. S9**, the thermal dependence of the charge carrier mobility showed a similar thermal hysteresis loop to that observed while exciting the laser at 530 nm. However, the hysteresis loop is expected to be inversed in this case as the amount of photogenerated carriers should increase upon the thermally-induced LS-HS transition at λ = 830 nm.[31] This observation leads to the conjecture that the dominant TRMC signal comes from the LS absorption band tail remaining at 830 nm (where $F_A$ is equal to 27 %) rather than from the corresponding maximum of the HS absorption band. Consequently, we could not quantitavely compare the mobilities in both spin-states.

**Temperature dependence of the charge carrier mobility from 77 K up to 360 K**

To get more insight into the charge transport mechanism in the LS state, we extended the TRMC measurements from liquid nitrogen temperature up to 360 K. **Figure 4** presents the natural logarithm values of the maximum TRMC signal against inverse temperature. A strong monotonic decrease of the mobility with temperature is found. Furthermore, two different regimes are found above and below a transition temperature of ca. 230 K. We extract the activation energies ($E_a$) using the following equation, above and below $T_R$, respectively:

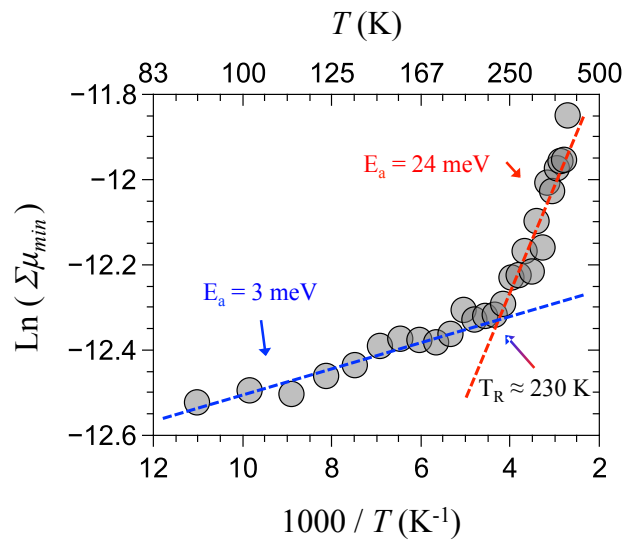

**Fig. 4** Natural logarithm of the maximum TRMC signal after laser pulse exciting at 530 nm versus reciprocal temperature. Two activation energies are deduced from an Arrhenius fit as presented by the dashed lines.

$$\Sigma\mu_{min} \propto \exp\left(-\frac{E_a}{k_B T}\right). \quad (3)$$

The high-temperature range for $T > T_R$ exhibits thermal activation with a small activation energy of $E_a$= 24 ± 2 meV. In contrast, the mobility at the lower temperature range, for $T < T_R$ can be viewed as activation-less. Indeed, the extracted activation energy of 2.9 ± 0.2 meV corresponding to ca. 34 K remains lower than the temperature at which it was measured. Similar activation energy values were obtained upon re-heating this sample to room temperature. Importantly, the temperature dependence of the empty quartz holder is distinctly different (see **Fig. S10**, Supporting Information).

It is interesting to note that the activation energy of 24 meV extracted near room temperature is more than one order of magnitude lower than that reported in the literature both in DC and AC modes for the same family of SCO compounds. Indeed, from DC transport measurements above room temperature the values extracted were 340-520 meV[32] and 417-1085 meV.[33] In the same conditions, AC transport measurements led to energy barriers of 500 and 200 meV at frequencies of 10 Hz and 1 MHz, respectively.[22] To explain these differences one may consider that TRMC is sensitive to carrier motion within the particles, i.e., intra-particle movement, while at MHz frequencies the charge carriers are more likely forced to cross several grain boundaries and/or interact stronger with impurities.[34] The difference in energy barrier can thus be understood from the length scales probed in the experiments: the lower limit of the charge carrier diffusion $x_{diff}$ inside the NPs occurring during one-half oscillating microwave cycle can be evaluated as:[30]

$$x_{diff} = \left(\frac{\Sigma\mu_{min} \cdot k_B \cdot T}{f \cdot e}\right)^{1/2}, \quad (4)$$

where $k_B$ is Boltzmann's constant, $T$ is the temperature, $f$ is the electric field frequency (9 GHz) and $e$ is the charge of an electron. This expression is valid for low microwave electric field strengths[34] ($\leq 100$ V.cm$^{-1}$). The diffusional motion falls in a range of 2 nm at 100 K up to 5 nm at 360 K. It represents a local displacement of 1-3 or 3-7 repeat unit cells respectively along the inter-chain $a$ and intra-chain $b$ axes of the crystal packing of [Fe(Htrz)$_2$(trz)](BF$_4$).[35]

Further insight can be obtained by considering the half-life times, defined as the time needed for the TRMC signal to decay to half of its maximum. Importantly, half-life times significantly vary with temperature as shown in **Fig. 5** while, as demonstrated in **Fig. S10** in Supporting Information, the reference measurements showed no temperature dependence of the decay time kinetics of the TRMC signal. **Figure 6** displays TRMC decay kinetics evaluated over the whole temperature range via a stretched exponential function up to the ms-range. Half-life times of *ca.* 200 µs take place at high temperature and decrease by one order of magnitude near liquid nitrogen temperature. These values are beyond the upper time window of the system of *ca.* 100 µs and suggest an efficient spatial separation of electrons and holes[35] especially at high temperatures. Besides, these TRMC signals follow two temperature-dependent exponential decays with a transition temperature $T_{R\ (decay)}$ of *ca.* 225 K, close to $T_R$ (*ca.* 230 K), as aforementioned. The observed decrease of the half-life times with temperature is in contrast to the temperature dependence reported in organic solar cell compounds, where higher mobilities lead usually to faster decays.[37] Therefore, we would like to stress that for these SCO compounds, the decay kinetics is not correlated to the charge carrier mobilities.

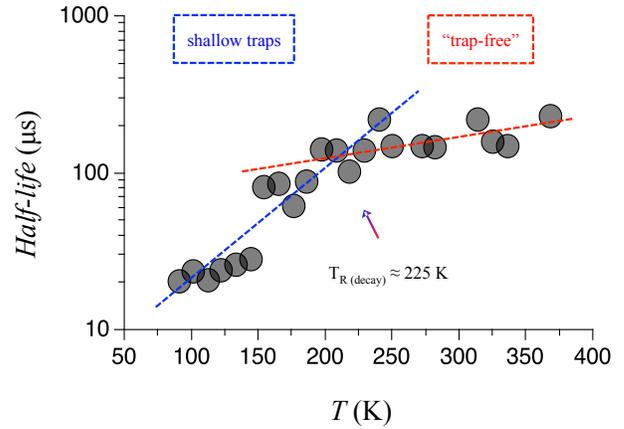

**Fig. 6** Half-life times, extracted from exponential fits of the TRMC signals shown in Fig. 5, as a function of temperature. Measurements were obtained with a 530 nm pump fluence of 2.45×10$^{15}$ photons per pulse. The dashed lines are drawn as a guide for the eyes.

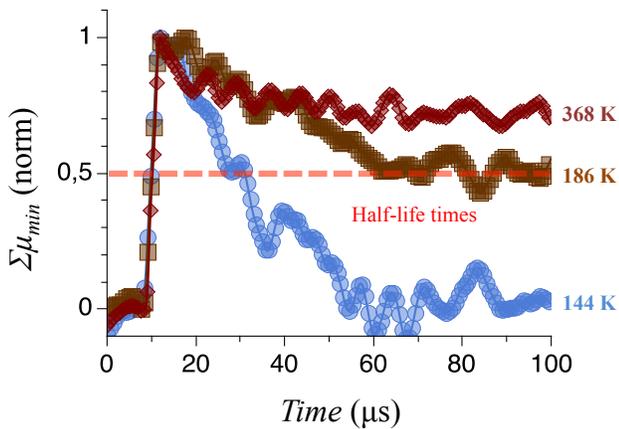

**Fig. 5** Normalized mobility decay dynamics versus time for different temperatures obtained after laser pulses exciting at 530 nm. Half-life times are plotted in Fig. 6 for the whole temperature range.

## Discussion

As $T_R$ is very close to $T_{R\ (decay)}$, it is tempting to relate the two temperature-dependent regimes of the mobility and the decay kinetics of the charge carriers above-mentioned. At low temperatures, below $T_R$, the small activation energy of 3 meV associated to short lifetimes of the charge carriers suggests that transport is limited by shallow trap states, away from a few $k_BT$ from the HOMO levels of the SCO NPs.[38] In other words, the averaged trapping time of charge carriers at shallow traps remains much longer than the average time of diffusive motion between the trapping events, which explains the low mobilities and short lifetimes observed at low temperatures. Above $T_R$, these trapped states are thermally released yielding a thermally activated hopping regime, where longer hops increase the mobility but, in counterpart, also the barrier energy. The activation energy could originate from inherent factors, such as molecular packing, dynamic disorder of the lattice (non polaronic in nature), polaronic self-localization, and competition between coherent and hopping motion of polarons.[39]

Efficient coupling of the charge carriers to the lattice vibrations of quasi-organic SCO compounds is expected, and even more pronounced for those exhibiting a wide memory effect, as strong cooperative interactions via electron-phonon coupling between molecules changing spin-states are required. For this reason, the observed thermally activated conductivity at higher temperatures in SCO compounds has been suggested to arise from a hopping polaron transport mechanism.[32] In other words, the charge carrier wave functions being localized on a few molecular units in our SCO compound could be of polaronic nature and would fall into different type of polaron formation depending on the electron hopping energy and the electron-phonon coupling,[40] which is specific for each SCO compound.

However, disorder is inevitably present in crystalline SCO compounds and could also contribute to the transport properties above $T_R$. In this context, diverse contributions of disorder should be considered: disorder has a time-dependent dynamic and/or a time-independent static origin. Dynamical disorder, present even for defectless ideal crystals, not only encompasses local polaronic carriers but also non-local lattice vibrations along and/or between interacting [Fe(Htrz)$_2$(trz)]$_n$ chains. It is important to realize that the small sizes of these NPs and the presence of the silica shell can also drastically modify the intrinsic phononic density of states of the NPs, as underlined recently.[41] On the other hand, static disorder of lattice vibrations may originate from both structural defects and impurities. For this SCO compound one can envision several types of chemical and physical defects specially in the vicinity of the SCO/silica interface such as 1D chain breaks, ring torsions of the triazoles leading to conformational torsions, bending and elongation of the 1D Fe$^{II}$ chains, but also a global loose of the orthorhombic crystal packing.[42] Consequently, static disorder results in erratic inter- and/or intra-molecular hopping integrals between sites causing variations of the kinetic energy of carrier hopping hindering high carrier mobilities as observed in this work.

It is important to mention that usually holes and electrons differ strongly in their sensitivity to be trapped at defects or impurities, or to interact with vibrational modes. However, the TRMC technique does not allow to distinguish between holes and electrons in nature. In the search of a comprehensive examination of the transport properties of SCO nanomaterials, a complementary direct-current dc method such as time-of-flight (TOF) combined to TRMC measurements could unveil the genuine nature of the charge carrier and give an estimation of the optical quantum yield ($\phi$). In further experiments the temperature and electric field dependence offered by the use of a TOF technique should shed light on the origin of the charge carrier trapping mechanism (polarons and/or disorder).

Besides, the use of a colour insensitive method such as electron pulse-radiolysis TRMC[43] might allow to compare the intrinsic mobilities of charge carriers in both spin-states. This work is in progress.

As non-spherical (SCO) objects can be manipulated to align their major axis in the direction of the electrical field,[16] we envision that tuning the polarization of a GHz-THz beam in-plane and out-of-plane could bring unassessed anisotropy transport (*i.e.,* comparison between intra and inter Fe(II) chain transport), which remains a fundamental question to tackle.

*Conclusion*

In conclusion, we have presented a contactless time-resolved microwave conductivity (TRMC) study on spin-crossover (SCO) compounds that offers an alternative to dielectric techniques and DC/AC electrical measurements performed between electrodes. Thanks to its GHz dynamics microwave probe, TRMC probes nanoscale charge transport of mobile photo-generated charge carriers in ~100 nm size hybrid SCO@SiO$_2$ NPs of [Fe(Htrz)$_2$(trz)](BF$_4$) compound. We have determined a lower limit of the local charge carrier mobility in the LS state, ranging from 5 up to 9×10$^{-6}$ cm$^2$.V$^{-1}$.s$^{-1}$ respectively from room temperature up to 360 K. TRMC measurements from 77 K up to 360 K, have shown two well-defined regimes in the mobility and the half-life times, which possess similar transition temperatures in a range of 225-250 K. We propose that these two transport regimes in the LS state can be rationalized in terms of a transition where charge carriers are tunnelling between shallow traps, towards a trap-free hopping regime thermally-activated above 225-250 K where longer hops increase the mobility but at the same time also the barrier energy.

**ACKNOWLEDGMENT**

The present work has been funded by the EU (COST Action CA15128 MOLSPIN and project COSMICS 766726), the Spanish MINECO (Unit of excellence "María de Maeztu" MDM-2015-0538 and project MAT2017-89993-R) and the Generalitat Valenciana (Prometeo Program of excellence). M.G.-M. thanks Spanish MICINN for a Juan de la Cierva-Incorporation grant. We thank Prof. Dr. Laurens Siebbeles and Dr. ir. T.J. Savenije for fruitful discussions.

**Experimental details:**

**Material synthesis**

**SiO$_2$@SCO NPs synthesis**

The hybrid [Fe(Htrz)$_2$(trz)](BF$_4$)@SiO$_2$ nanoparticles (NPs) have been synthesized following the method developed by Herrera and Colacio and co-workers.[1] An aqueous solution of Fe(BF$_4$)$_2$·6H$_2$O (0.5 mL, 1.25 M) with *ca*. 2 mg of ascorbic acid and tetraethyl orthosilicate (TEOS) (0.1 mL) is added to a previously prepared solution of Triton X-100 (1.8 ml, ω= 9), hexanol (1.8 mL) and cyclohexane (7.5 mL). The obtained mixture is stirred at room temperature for 15 min, in order to get a stable microemulsion. Similarly, a solution of 1,2,4-1H-Triazole (0.5 mL, 3.75 M) ligand and TEOS (0.1 mL) is added to a second solution of Triton X-100, hexanol and cyclohexane. Both microemulsions are combined and left to react for 24 h to ensure micellar exchange. Finally, NPs are obtained by precipitation upon addition of acetone to destabilize the microemulsion and then collected by centrifugation (12000 rpm, 10 min), followed by 4 cycles of washing with ethanol, in order to remove the excess of surfactant, and one with acetone. The powdered samples are dried at 70 °C for 2 h prior to use.

**Characterization**

**The optical spectrum** of the SCO@SiO$_2$ NPs in a form of powder placed inside the quartz cuvet used for TRMC measurements was recorded using a spectrophotometer (Ocean Optics USB2000+).

**Diffuse reflectance spectra** were recorded on powder samples in the 350–800 nm range using a CARY 5G UV-Vis-NIR spectrophotometer equipped with an integrating sphere attachment. Reflectances of the sample and the integration sphere used as a high-absorbing reference, were measured under the same conditions. The spectrum resolution was 1 nm and the scan rate 150 nm/min.

**NPs size and shape were characterized by High-resolution transmission electron microscopy (HR-TEM)** using a TECNAI G$^2$F20 S-TWIN HR microscope operating at 200 kV. Sample preparation was done placing a drop of the colloidal suspension containing the SCO@SiO$_2$ NPs onto a carbon coated copper grid. NPs size distribution was also analysed in suspension by dynamic light scattering (DLS) using a Zetasizer ZS (Malvern Instrument, UK) and then determined by "manual counting" using image-J software.

**Energy dispersive X-ray spectroscopy (EDX) elemental mapping images** of the NPs were collected on a Philips XL-30 ESEM microscope.

**Magnetic susceptibility measurements** were performed on powdered samples of [Fe(Htrz)$_2$(trz)](BF$_4$)@SiO$_2$ with a Quantum Design MPMS-XL-5 SQUID susceptometer in the temperature range 300-400 K under an applied magnetic field of 0.1 T (heating/cooling rate of 1 K.min$^{-1}$). The susceptibility data were corrected from the diamagnetic contributions as deduced by using Pascal's constant tables.

**TRMC**

A description of the TRMC setup was already detailed elsewhere. Briefly, the sample is photo-excited with a short (3 ns) linearly polarized laser pulse from an optical parametric oscillator pumped at 355 nm with the third harmonic of a Q-switched Nd:YAG laser (Vibrant II, Opotek). The product of the charge carrier generation quantum yield and the sum of the electron and hole mobilities, ϕ.Σμ, is usually determined from the maximum in the time-dependent photoconductivity.

It is important to note that local displacements of charge carriers after a laser pulse are significantly larger than the charge drift distances induced from the AC field perturbation that we estimated inferior to 2-5 pm.[4]

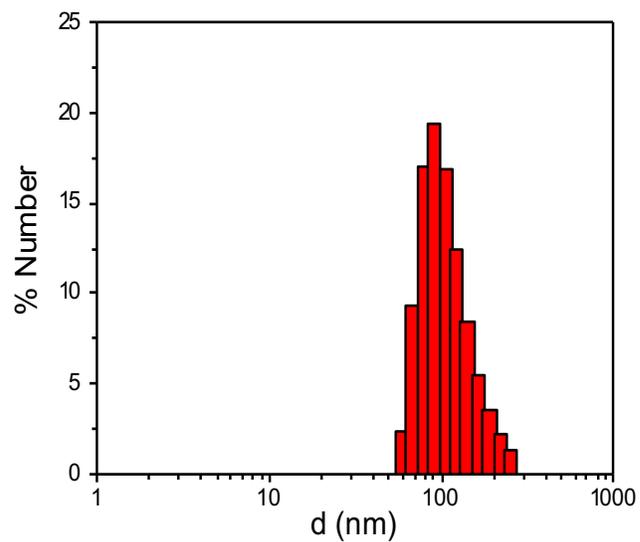 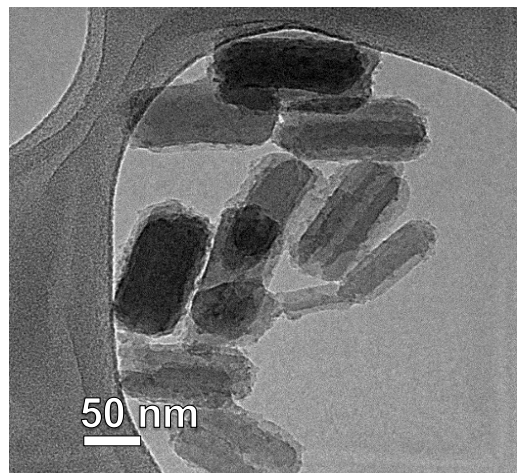

Figure S1 (left) Number-based size distribution of a core-shell NP colloidal suspension obtained by DLS analysis displaying sizes of about 100 nm. (right) Transmission electron microscopy images of the same NPs deposited by drop casting on a carbon coated copper grid showing the hybrid SCO-core and much less dense $SiO_2$-shell structure. Scale bars of 50 nm.

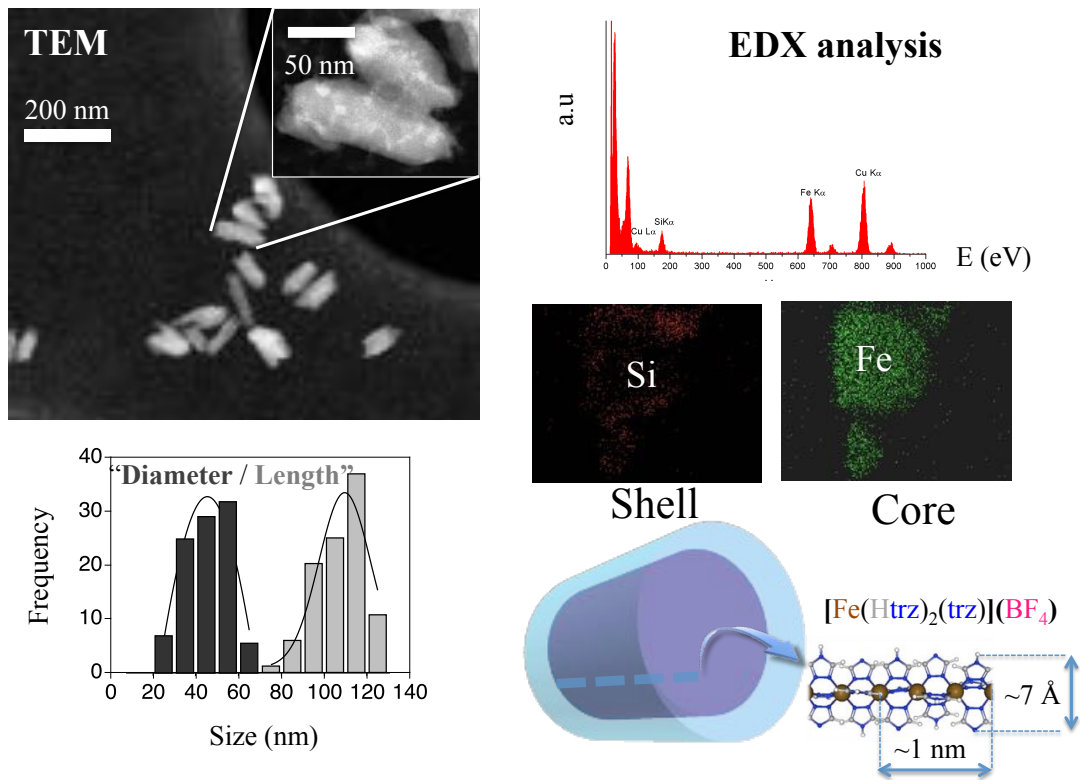

Figure S2 (top left) Representative high-angle annular dark field scanning transmission electron microscopy image of hybrid spin-crossover@SiO$_2$ nanoparticles. (down) Corresponding "Diameter" / Length distribution of the nanoparticles. (top right) Energy dispersive X-ray spectroscopy (EDS) elemental mapping images of the NPs. (down right) schematic representation of the core-shell nanoparticle and chemical structure of the 1-D coordination polymer of the SCO core.

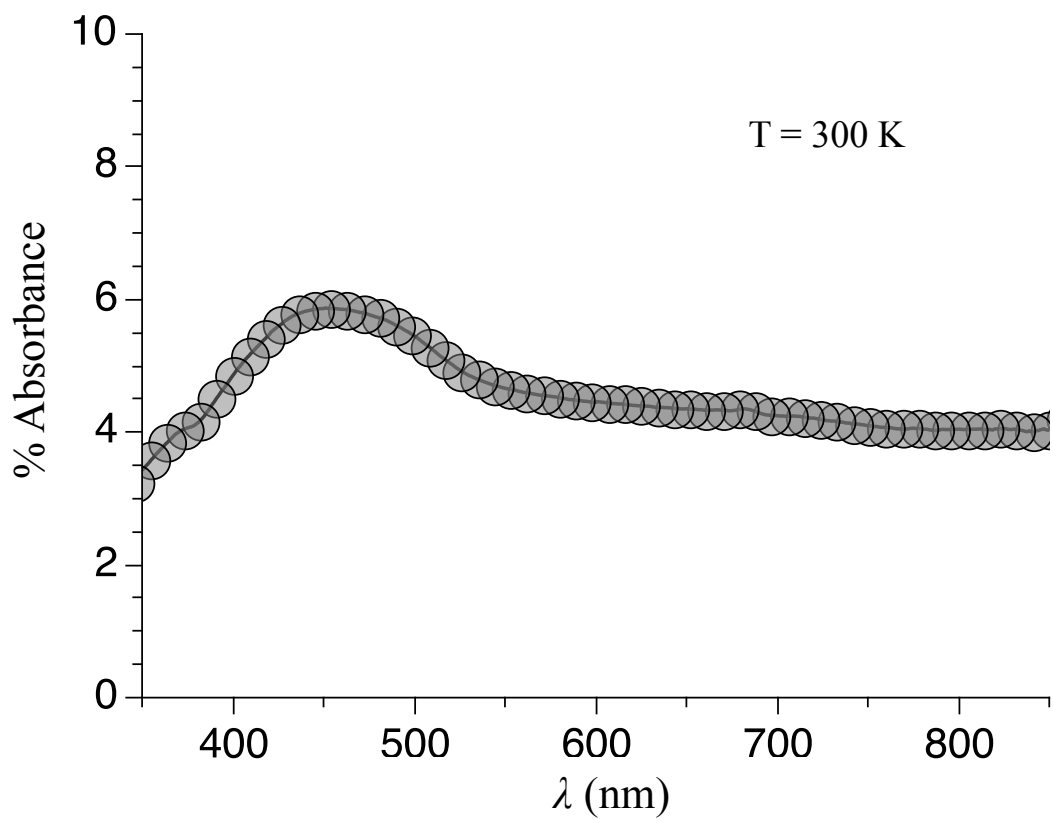

**Figure S3** Absorbance spectrum recorded at room temperature of a powder reference sample made of pure silica placed inside the quartz cuvet employed for TRMC measurements. At maximum of *ca.* 6 % and only of *ca.* 5 and 4 % of light is adsorbed at 530 nm (*i.e.* when the NP core is in the low-spin state) and 830 nm (*i.e.* when the NP core is in the high-spin state), respectively. One can conclude that the light absorption of the silica shells of the hybrid nanoparticles is negligible.

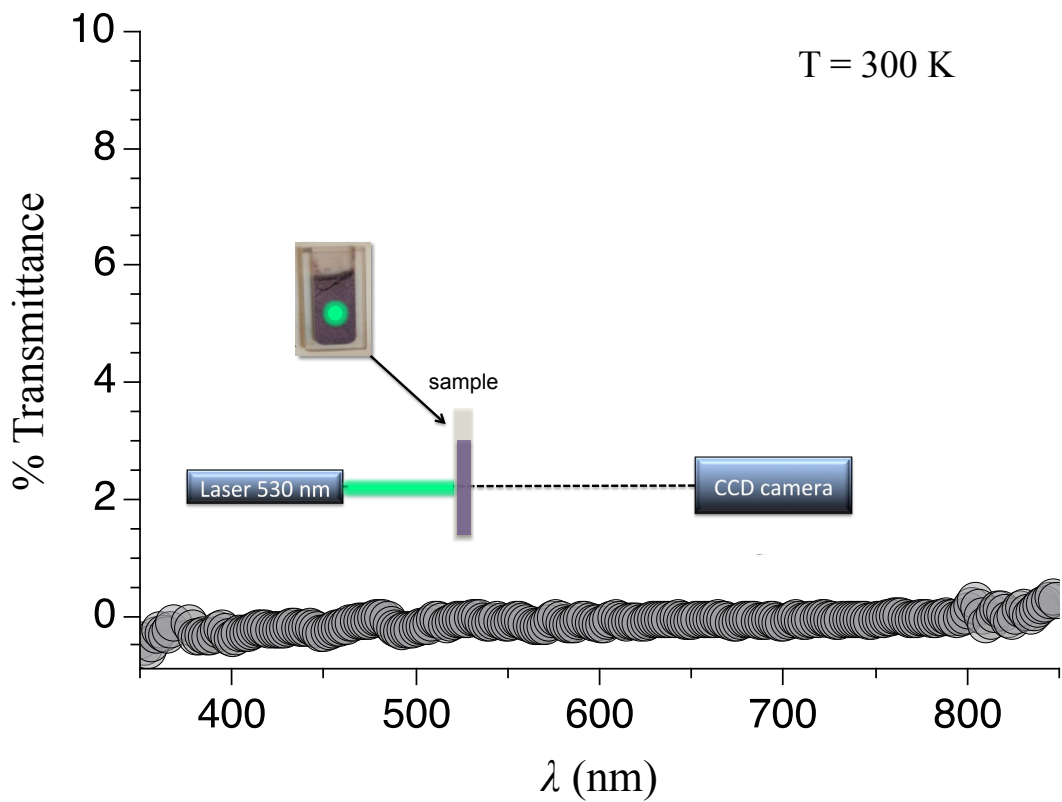

**Figure S4** Incident laser light fraction transmitted through a sample made of a quartz cuvet filled in with a powder of hybrid SCO@SiO$_2$ NPs. The beam was focused at the center of the sample and almost no light is transmitted through the sample in the wavelength range of 300-900 nm.

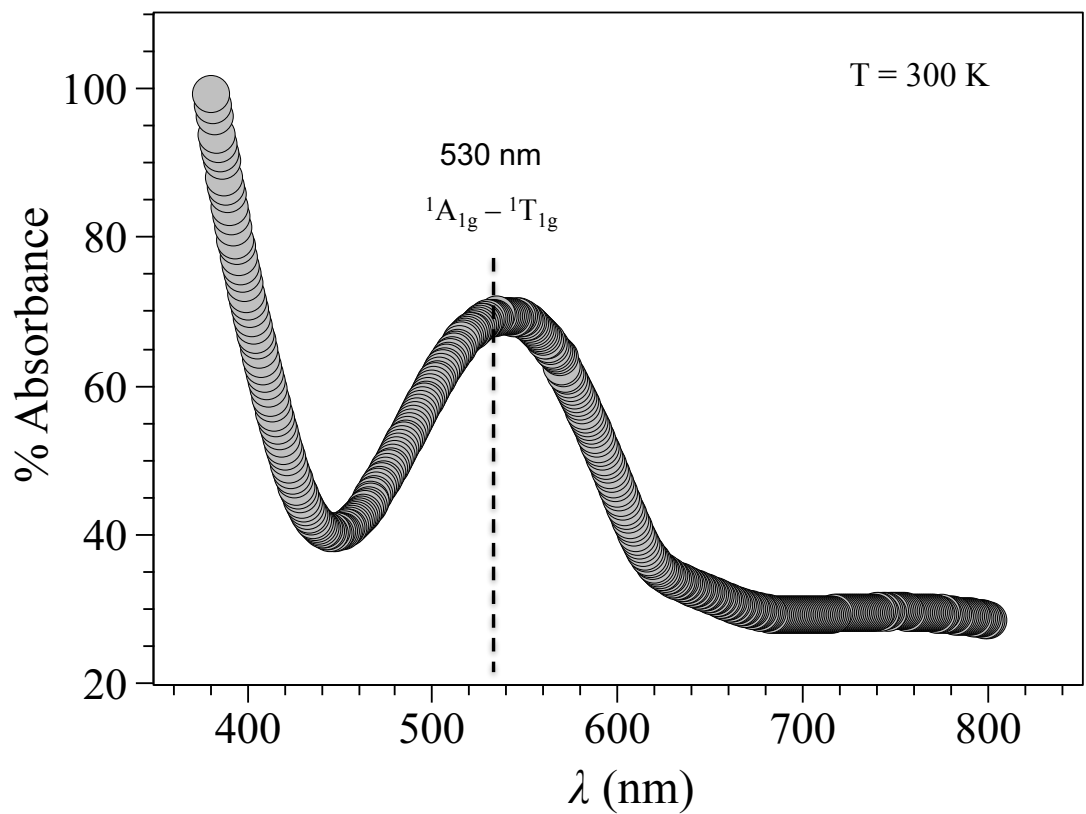

**Figure S5** Diffuse reflectance spectrum recorded at room temperature of the hybrid SCO@SiO$_2$ NPs. One can observe the well-known absorption band centred at 530 nm corresponding to one of the two spin-allowed ligand-field transitions $^1A_1$ - $^1T_1$ in the LS state.

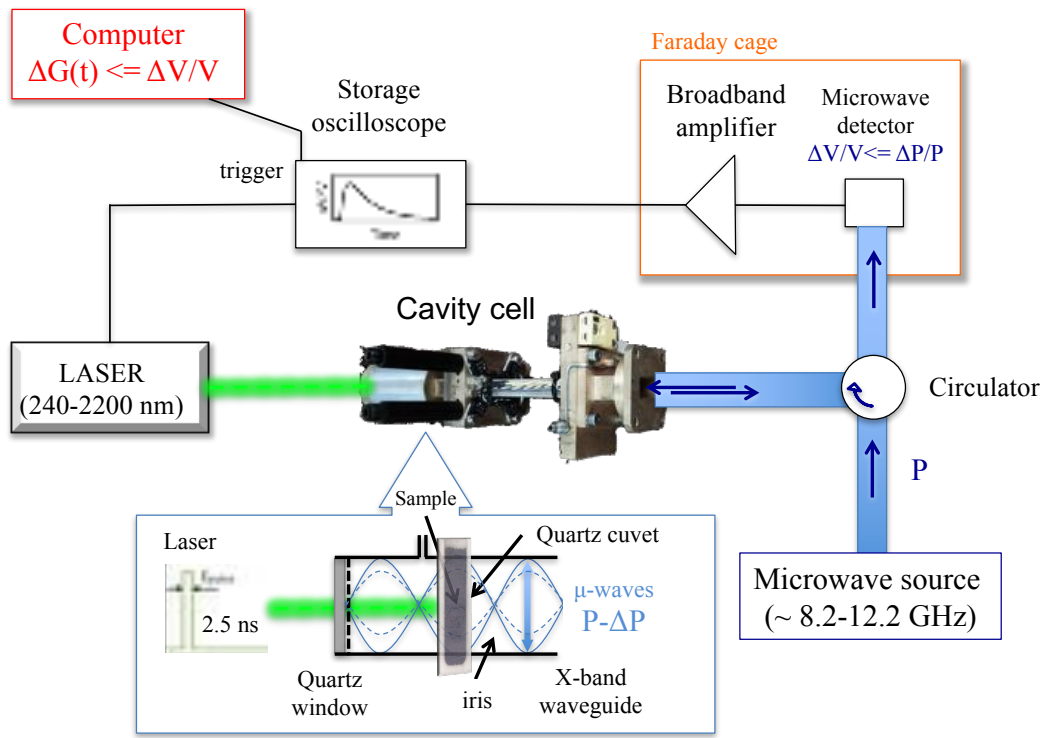

**Figure S6** Schematic representation of the time-resolved microwave photoconductance setup. The interior of the cavity cell is sketched at the bottom of this figure, in which one can observe the quartz cuvet containing a powder of SCO@SiO$_2$ NPs. The physical principle underlying the TRMC technique is the proportional relationship of the change in conductance with the attenuation of microwaves propagating through a weakly conducting medium (**P − ΔP**). Adapted from Savenije et al.[3]

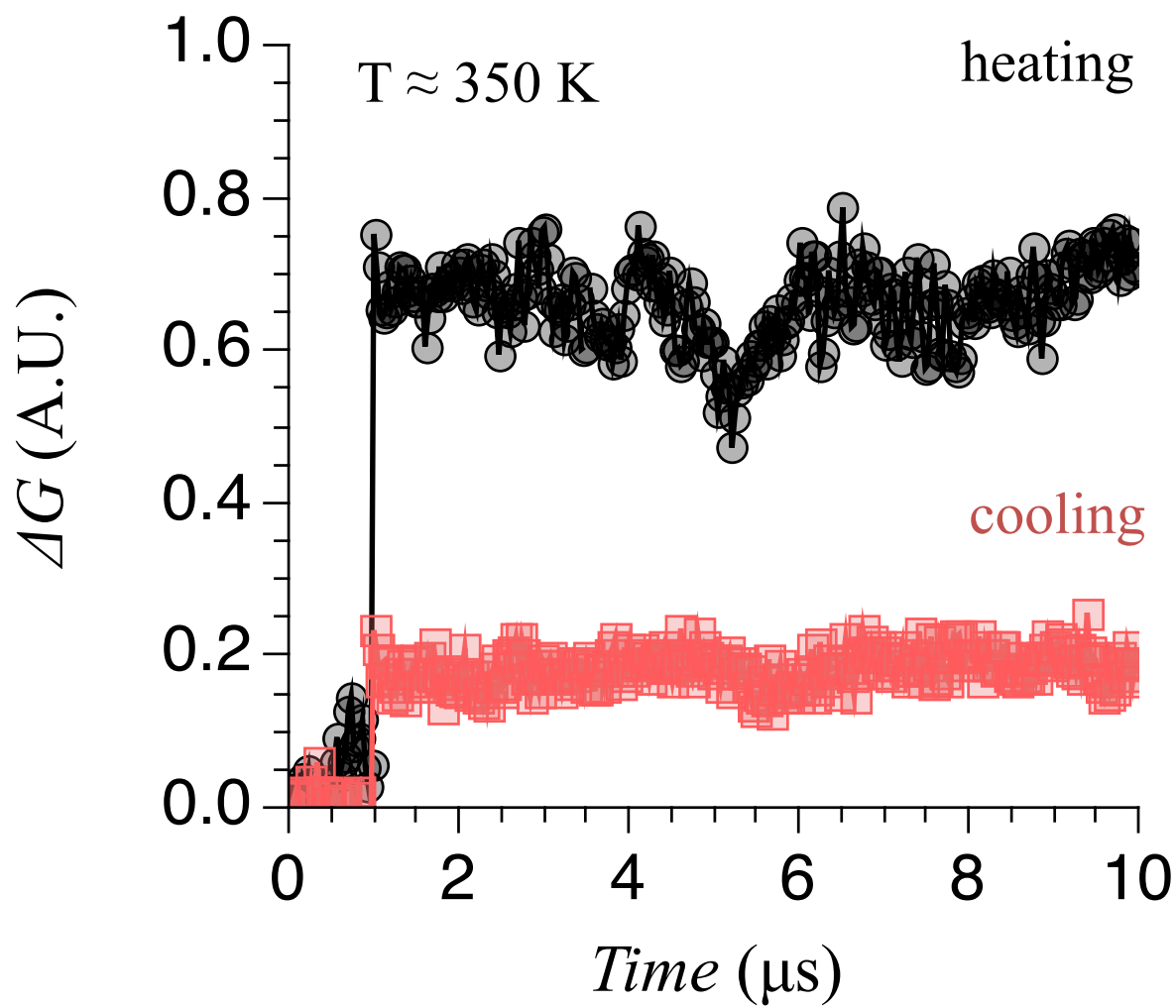

**Figure S7** Black and red curves correspond respectively to TRMC signal change after laser pulse exciting at 530 nm of a spin-crossover nanoparticle powder sample in the heating and cooling mode at T = 350 K. Measurements were averaged over 1000 laser pulses with a pump fluence of $2.45\times10^{15}$ photons per pulse. The dark conductance has been substrated for all data points.

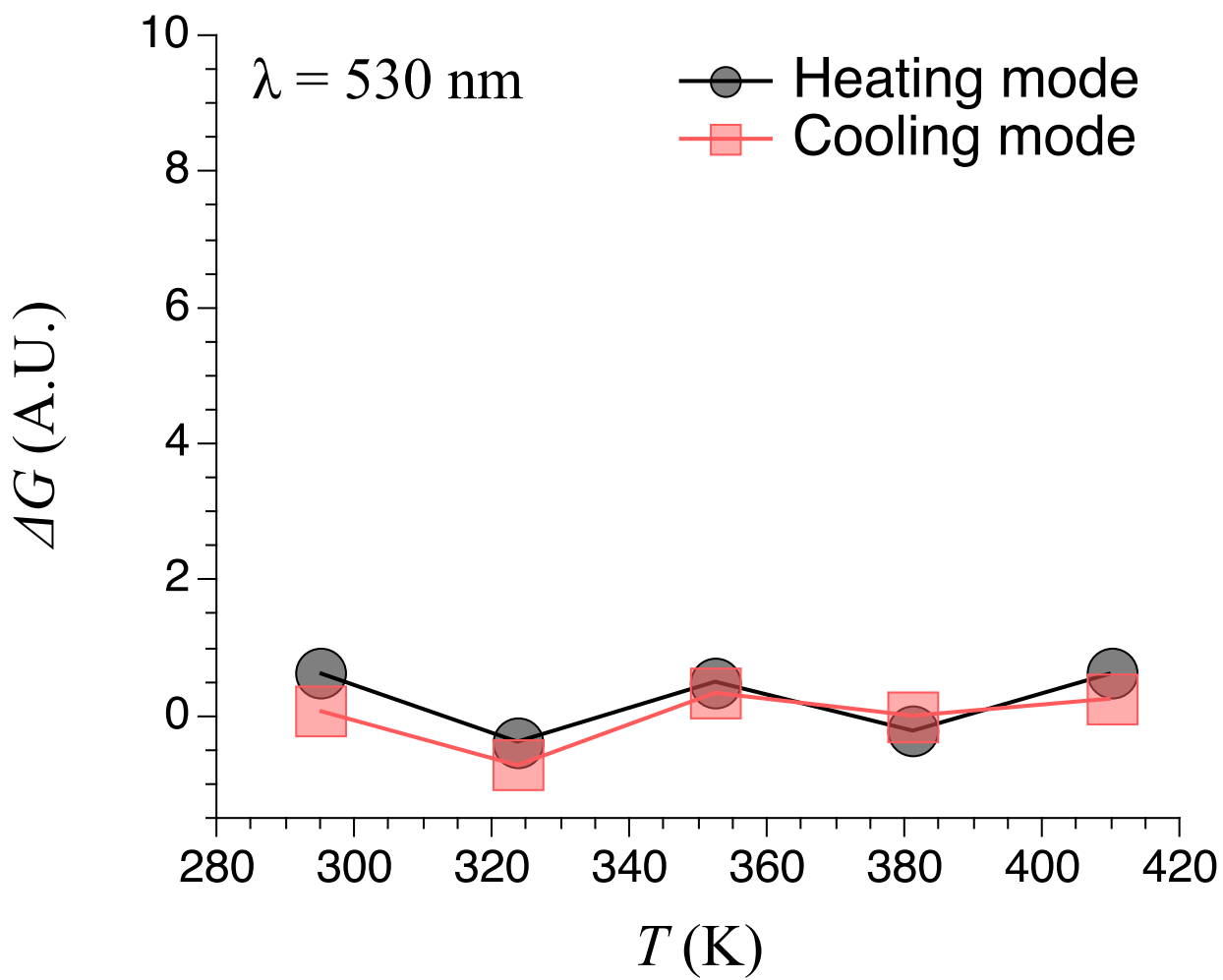

**Figure S8** Maximum observed TRMC signals as a function of temperature of the referene sample (empty quartz cuvet). The heating and cooling modes corresponding respectively to the red and black curves didn't exhibit any temperature dependence. Measurements were averaged over 1000 laser pulses and obtained with a 530 nm pump fluence of 2.45×10$^{15}$ photons per pulse. The dark conductance has been substrated for all data points.

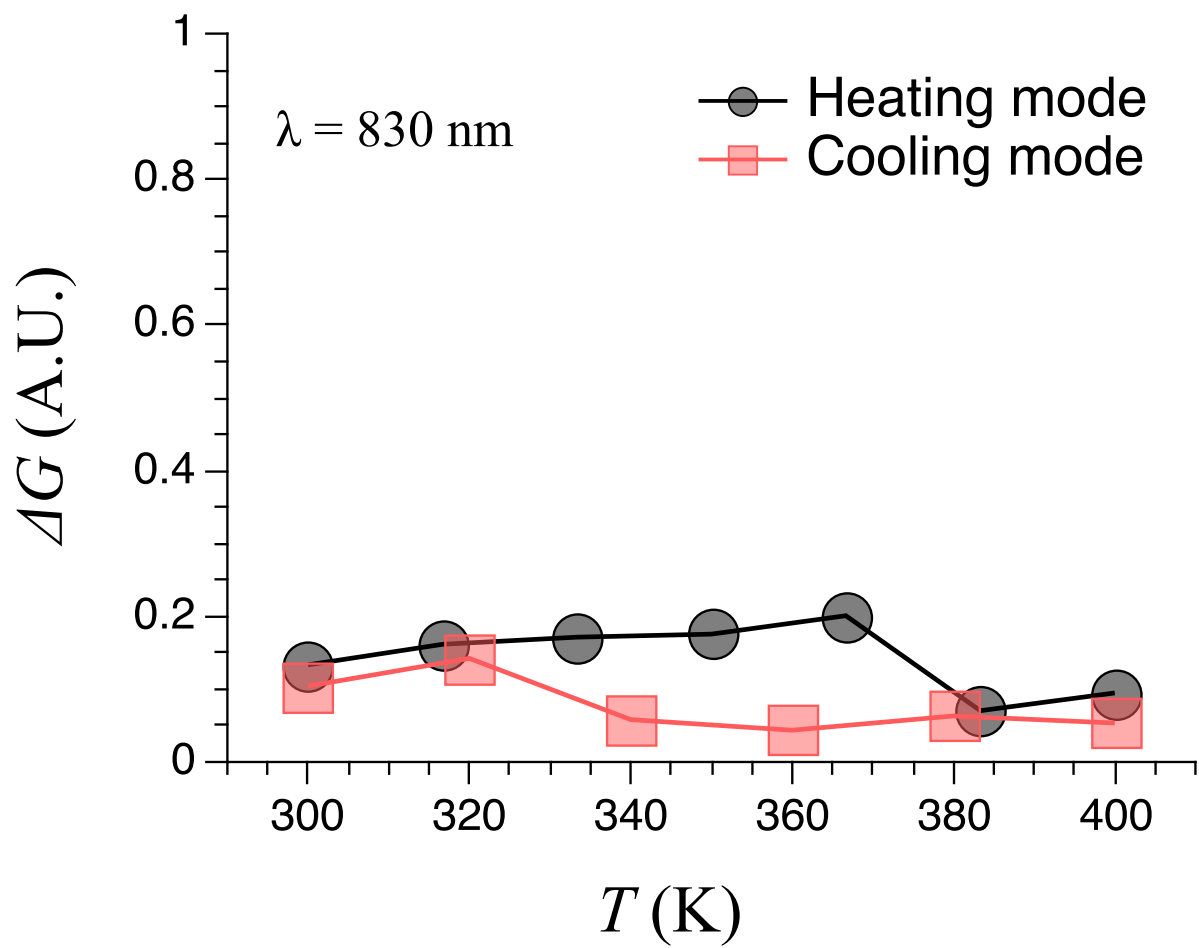

**Figure S9** Maximum conductance change against temperature (300-400 K). Measurements were averaged over 1000 laser pulses and obtained with a 830 nm pump fluence of $2.45\times10^{15}$ photons per pulse. The dark conductance has been substrated for all data points.

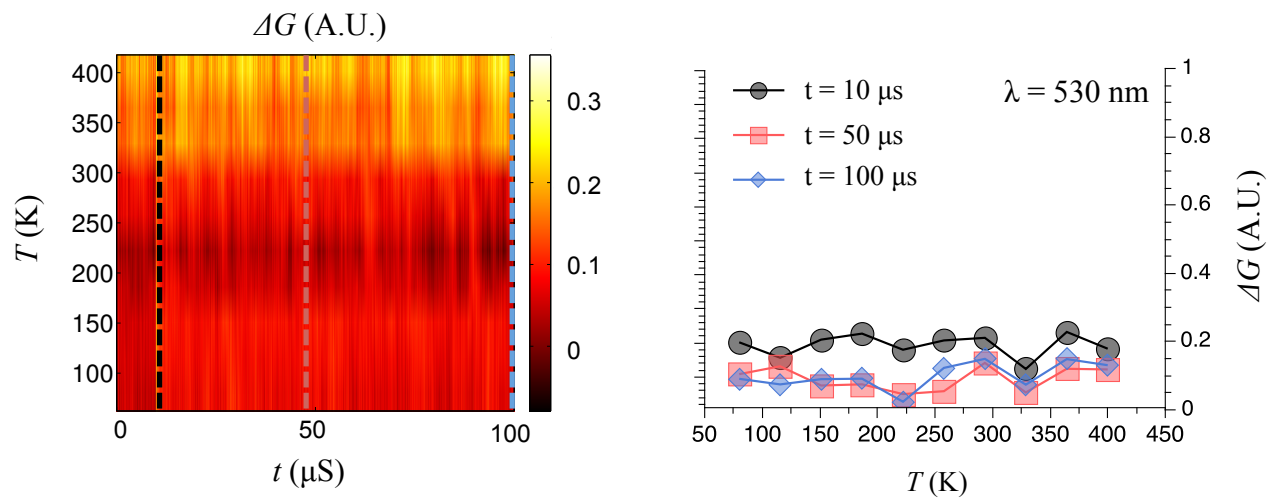

**Figure S10** Left: 2D-TRMC signals as a function of time for the referene sample, *i.e.* empty quartz cuvet, recorded from liquid nitrogen temperature up to 400 K. Right: Corresponding maximum TRMC signals plotted at different time after the laser pulse, namely 10, 50 and 100 µs, demonstrating that the reference of the study didn't show any temperature dependence of the signal decay kinetics and nor of the mobility. Measurements were averaged over 1000 laser pulses and obtained with a 530 nm pump fluence of $2.45\times10^{15}$ photons per pulse. The dark conductance has been substrated for all data points.

One can note that the hysteresis loop width of ca. 25 K is close, but nevertheless smaller than the one of ca. 40 K shown in **Fig. 2 (**see in the main text) acquired using a SQUID magnetometer. We ascribe this reduction to the wavelength excitation used for TRMC measurements, which corresponds to a penetration depth within the sample of ~100 nm[5,6] *i.e.,* a monolayer of NP is probed whereas the full NP size population is sensed using a SQUID magnetometer. Alternatively, one can also wonder whether the closening of the hysteresis loop could originate from a laser-induced process, *i.e.,* either photo-induced and/or thermally-driven processes. Indeed, above a certain laser energy density threshold it is known that the spin-transition can be triggered within the thermal hysteresis loop reducing its loop width. In our work however, we employed a 2 mJ.cm$^{-2}$ laser power, which remains below the values reported in the literature required to induce the LS to HS transition for this family of SCO material by means of both a continuous[7] or a pulsed laser.[8,9]